\documentclass[10pt,A4]{article}
\usepackage{graphicx}
\usepackage{hyperref}
\usepackage{tikz}
\def\R2Lurl#1#2{\mbox{\href{#1}{\tt #2}}}
 \usepackage[]{hyperref}
\usepackage{amsmath, amsthm, amssymb}

\newcommand{\csab}{$\cos^{2}(\alpha - \beta)$}
\newcommand{\cscb}{$\cos^{2}(\beta - \gamma)$}

\newcommand{\figi}{\begin{center}
\begin{tikzpicture}
\begin{scope}[scale=1.2]
\draw[color=black,ultra thin] (-3.2,-3)--(-3.2,3.2)--(3.2,3.2)--(3.2,-3)--(-3.2,-3); 
\draw[color=black,ultra thin] (-2,-2)--(1.5,1.5); 
\draw[color=black,ultra thin] (2,-2)--(-1.5,1.5); 
\draw[color=black, ->] (-2.5,-2)--(-2.5,0.25); 
\draw[color=gray,ultra thick] (2.25,1.75)--(1.75,1.25)--(1.25,1.75)--(1.75,2.25)--(2.25,1.75)--(1.75,1.25);
\draw[color=gray,thick] (1.75,1.4)--(1.75,2.1);
\draw[color=blue,densely dashed] (2,2)--(2.5,2.5);
\draw[color=blue,densely dashed] (1.5,2)--(1.0,2.5);
\draw[color=gray,ultra thick] (-2.25,1.75)--(-1.75,1.25)--(-1.25,1.75)--(-1.75,2.25)--(-2.25,1.75)--(-1.75,1.25);
\draw[color=gray,thick] (-1.75,1.4)--(-1.75,2.1);
\draw[color=blue,densely dashed] (-1.5,2)--(-1,2.5);
\draw[color=blue,densely dashed] (-2,2)--(-2.5,2.5);
\draw[color=blue,thick, ->] (0,0)--(1.5,1.5); 
\draw[color=blue,thick, ->] (0,0)--(-1.5,1.5); 

\draw (0.,-3.5) node {{\small\emph{Figure 1: EPRB (with photons).}}};
\draw (0.75,1) node  [rotate=45] {{\scriptsize Photon}};
\draw (-1.25,-1.04) node  [rotate=45] {{\scriptsize Lightcone}};
\draw (-2.7,-1) node  [rotate=90] {{\footnotesize Time}};
\draw (0.,-0.4) node {M};
\draw (2.5,2.7) node {{\footnotesize $\textup{\textbf{B}}=1$}};
\draw (-2.5,2.7) node {{\footnotesize $\textup{\textbf{A}}=1$}};
\draw (1,2.7) node {{\footnotesize $\textup{\textbf{B}}=0$}};
\draw (-1,2.7) node {{\footnotesize $\textup{\textbf{A}}=0$}};
\draw (2,0.8) node {{\scriptsize\parbox{2cm}{{\scriptsize Polarizing~cube\\ set at angle $\beta$}}}};
\draw (-2,0.8) node {{\scriptsize\parbox{2cm}{Polarizing~cube\\ set at angle $\alpha$}}};
\draw (0,-2.35) node {{\footnotesize $Pr(\textup{\textbf{A}}=\textup{\textbf{B}}) =\cos^{2}(\alpha -\beta)$}};
\end{scope}
\end{tikzpicture}
\end{center}}

\newcommand{\figii}{\begin{center}
\begin{tikzpicture}
\begin{scope}[scale=1.2]
\draw[color=black,ultra thin] (-3.2,-3.6)--(-3.2,3.2)--(3.2,3.2)--(3.2,-3.6)--(-3.2,-3.6); 
\draw[color=black,ultra thin] (-2.5,-2.5)--(1.5,1.5); 
\draw[color=black,ultra thin] (1.5,-1.5)--(-2.5,2.5); 
\draw[color=black, ->] (-2.5,-1.5)--(-2.5,1.5); 
\draw[color=gray,ultra thick] (2.25,1.75)--(1.75,1.25)--(1.25,1.75)--(1.75,2.25)--(2.25,1.75)--(1.75,1.25);
\draw[color=gray,thick] (1.75,1.4)--(1.75,2.1);

\draw[color=blue,densely dashed] (2,2)--(2.5,2.5);
\draw[color=blue,densely dashed] (1.5,2)--(1.0,2.5);

\draw[color=gray,ultra thick] (2.25,-1.75)--(1.75,-1.25)--(1.25,-1.75)--(1.75,-2.25)--(2.25,-1.75)--(1.75,-1.25);
\draw[color=gray,thick] (1.75,-1.4)--(1.75,-2.1);
\draw[color=blue,thick] (2,-2)--(2.5,-2.5);
\draw[color=gray] (1,-2.5)--(1.5,-2);
\draw (1,-2.7) node {{\footnotesize $\textup{\textbf{C}}=0$}}; 
\draw (2.5,-2.7) node {{\footnotesize $\textup{\textbf{C}}=1$}};
\draw (2.5,2.7) node {{\footnotesize $\textup{\textbf{B}}=1$}};
\draw (1,2.7) node {{\footnotesize $\textup{\textbf{B}}=0$}};

\draw[color=blue,thick, ->] (0,0)--(1.5,1.5); 
\draw[color=blue,thick] (1.5,-1.5)--(0,0); 

\draw[color=red,ultra thick] (0,-0.5)--(0,0.5); 

\draw (0.,-4.1) node {{\small\emph{Figure 2: Sideways EPRB (SEPRB).}}};
\draw (1,-0.77) node  [rotate=315] {{\scriptsize Photon}};
\draw (-1.25,-1.04) node  [rotate=45] {{\scriptsize Lightcone}};
\draw (-2.7,0) node  [rotate=90] {{\footnotesize Time}};
\draw (0.,-0.6) node {{\scriptsize Mirror}};

\draw (0.25,1.75) node {{\scriptsize\parbox{2cm}{{\scriptsize Polarizing cube\\ set at angle $\beta$}}}};
\draw (0.25,-1.75) node {{\scriptsize\parbox{2cm}{Polarizing cube\\ set at angle $\gamma$}}};

\draw (2.7,-2.4) node {{\small $\textup{S}$}};
\draw (-2.7,-2.4) node {{\small $\textup{T}$}};
\draw (0,-3.15) node {{\footnotesize $Pr(\textup{\textbf{B}}=\textup{\textbf{C}}) =\cos^{2}(\gamma -\beta)$}};

\end{scope}

\end{tikzpicture}
\end{center}}

\newcommand{\figiii}{\definecolor{orange}{rgb}{1,0.5,0}
\begin{center}
\begin{tikzpicture}
\begin{scope}[scale=1.2]

\shade[inner color=yellow!50!white!50, outer color=yellow!50!white!50] (0,-2.5) -- (0,0) -- (2.5,0) -- (2.5,-2.5) -- (0,-2.5);
\draw[color=black,ultra thin] (-3.2,-3.2)--(-3.2,3.2)--(3.2,3.2)--(3.2,-3.2)--(-3.2,-3.2); 
\draw[color=black,ultra thin] (-2.5,-2.5)--(1.5,1.5); 
\draw[color=black,ultra thin] (1.5,-1.5)--(-2.5,2.5); 
\draw[color=black, ->] (-2.5,-1.5)--(-2.5,1.5); 
\draw[color=orange, ultra thick, ->] (1.75,0.25) .. controls (1.75,0.75) .. (1.2,1); 
\draw[color=orange, ultra thick, ->] (1.75,-0.25) .. controls (1.75,-0.75) .. (1.2,-1); 
\draw[color=gray,ultra thick] (2.25,1.75)--(1.75,1.25)--(1.25,1.75)--(1.75,2.25)--(2.25,1.75)--(1.75,1.25);
\draw[color=gray,thick] (1.75,1.4)--(1.75,2.1);
\draw[color=blue,densely dashed] (2,2)--(2.5,2.5);
\draw[color=blue,densely dashed] (1.5,2)--(1.0,2.5);
\draw[color=gray,ultra thick] (2.25,-1.75)--(1.75,-1.25)--(1.25,-1.75)--(1.75,-2.25)--(2.25,-1.75)--(1.75,-1.25);
\draw[color=gray,thick] (1.75,-1.4)--(1.75,-2.1);

\draw[color=blue,thick] (2,-2)--(2.5,-2.5);
\draw[color=gray] (1,-2.5)--(1.5,-2);
\draw (1,-2.7) node {{\footnotesize $\textup{\textbf{C}}=0$}}; 
\draw (2.5,-2.7) node {{\footnotesize $\textup{\textbf{C}}=1$}};
\draw (2.5,2.7) node {{\footnotesize $\textup{\textbf{B}}=1$}};
\draw (1,2.7) node {{\footnotesize $\textup{\textbf{B}}=0$}};

\draw[color=blue,thick, ->] (0,0)--(1.5,1.5); 
\draw[color=blue,thick] (1.5,-1.5)--(0,0); 

\draw[color=red,ultra thick] (0,-0.5)--(0,0.5); 

\draw (0.,-3.8) node {{\small \parbox{9cm}{\begin{center}
\emph{Figure 4: The intuitive causal explanation  in SEPRB: a beable carries information about
the C setting to B.}
\end{center}
}}};
\draw (1.8,0) node {{\scriptsize C-beable:[$\gamma$]}};
\draw (-1.25,-1.04) node  [rotate=45] {{\scriptsize Lightcone}};
\draw (-2.7,0) node  [rotate=90] {{\footnotesize Time}};
\draw (0.,-0.6) node {{\scriptsize Mirror}};
\draw (0.25,1.75) node {{\scriptsize\parbox{2cm}{{\scriptsize Polarizing cube\\ set at angle $\beta$}}}};
\draw (0.25,-1.75) node {{\scriptsize\parbox{2cm}{Polarizing cube\\ set at angle $\gamma$}}};
\end{scope}

\end{tikzpicture}
\end{center}}

\newcommand{\figiv}{\begin{center}
\begin{tikzpicture}
\begin{scope}[scale=1.2]
\draw[color=black,ultra thin] (-3.2,-2.5)--(-3.2,3.2)--(3.2,3.2)--(3.2,-2.5)--(-3.2,-2.5); 
\draw[color=black,ultra thin] (-2,-2)--(1.5,1.5); 
\draw[color=black,ultra thin] (2,-2)--(-1.5,1.5); 
\draw[color=black, ->] (-2.5,-2)--(-2.5,0.25); 
\draw[color=orange, ultra thick, ->] (0.1,1.5) .. controls (0.22,0.9) .. (0.5,0.6); 
\draw[color=orange, ultra thick, ->] (-0.1,1.5) .. controls (-0.22,0.9) .. (-0.5,0.6); 
\draw (0, 1.7) node {{\scriptsize A-beable:[$\alpha$]}};
\draw[color=gray,ultra thick] (2.25,1.75)--(1.75,1.25)--(1.25,1.75)--(1.75,2.25)--(2.25,1.75)--(1.75,1.25);
\draw[color=gray,thick] (1.75,1.4)--(1.75,2.1);
\draw[color=blue,densely dashed] (2,2)--(2.5,2.5);
\draw[color=blue,densely dashed] (1.5,2)--(1.0,2.5);
\draw[color=gray,ultra thick] (-2.25,1.75)--(-1.75,1.25)--(-1.25,1.75)--(-1.75,2.25)--(-2.25,1.75)--(-1.75,1.25);
\draw[color=gray,thick] (-1.75,1.4)--(-1.75,2.1);
\draw[color=blue,densely dashed] (-1.5,2)--(-1,2.5);
\draw[color=blue,densely dashed] (-2,2)--(-2.5,2.5);
\draw[color=blue,thick, ->] (0,0)--(1.5,1.5); 
\draw[color=blue,thick, ->] (0,0)--(-1.5,1.5); 
\draw (0.,-3.1) node {{\small \parbox{9cm}{\begin{center}
\emph{Figure 5: The SERPB explanation applied to EPRB yields a model with a
beable that depends on the measurement setting $\alpha$.}
\end{center}
}}};

\draw (-2.7,-1) node  [rotate=90] {{\footnotesize Time}};
\draw (0.,-0.4) node {M};

\draw (2.5,2.7) node {{\footnotesize $\textup{\textbf{B}}=1$}};
\draw (-2.5,2.7) node {{\footnotesize $\textup{\textbf{A}}=1$}};
\draw (1,2.7) node {{\footnotesize $\textup{\textbf{B}}=0$}};
\draw (-1,2.7) node {{\footnotesize $\textup{\textbf{A}}=0$}};

\draw (2,0.8) node {{\scriptsize\parbox{2cm}{{\scriptsize Polarizing cube\\ set at angle $\beta$}}}};
\draw (-2,0.8) node {{\scriptsize\parbox{2cm}{Polarizing cube\\ set at angle $\alpha$}}};

\end{scope}
\end{tikzpicture}
\end{center}}

\newcommand{\figv}{\begin{center}
\begin{tikzpicture}
\begin{scope}[scale=1.2]

\draw[color=black,ultra thin] (-3.2,-2.5)--(-3.2,3.2)--(3.2,3.2)--(3.2,-2.5)--(-3.2,-2.5); 
\draw[color=black,ultra thin] (-2,-2)--(1.5,1.5); 
\draw[color=black,ultra thin] (2,-2)--(-1.5,1.5); 
\draw[color=black, ->] (-2.5,-2)--(-2.5,0.25); 
\draw[color=orange, ultra thick, ->] (0.1,1.5) .. controls (0.22,0.9) .. (0.5,0.6); 
\draw[color=orange, ultra thick, ->] (-0.1,1.5) .. controls (-0.22,0.9) .. (-0.5,0.6); 
\draw (0, 1.7) node {{\scriptsize A-beable:[$\alpha$]}};
\draw[color=orange, ultra thick, ->] (-0.1,-1.5) .. controls (0,-1) and (-1.2,-0.5) .. (-0.6,0.5); 
\draw[color=orange, ultra thick, ->] (0.1,-1.5) .. controls (0,-1) and (1.2,-0.5) .. (0.6,0.5);  
\draw (0, -1.7) node {{\scriptsize B-beable:[$\beta$]}};
\draw[color=gray,ultra thick] (2.25,1.75)--(1.75,1.25)--(1.25,1.75)--(1.75,2.25)--(2.25,1.75)--(1.75,1.25);
\draw[color=gray,thick] (1.75,1.4)--(1.75,2.1);
\draw[color=blue,densely dashed] (2,2)--(2.5,2.5);
\draw[color=blue,densely dashed] (1.5,2)--(1.0,2.5);
\draw[color=gray,ultra thick] (-2.25,1.75)--(-1.75,1.25)--(-1.25,1.75)--(-1.75,2.25)--(-2.25,1.75)--(-1.75,1.25);
\draw[color=gray,thick] (-1.75,1.4)--(-1.75,2.1);
\draw[color=blue,densely dashed] (-1.5,2)--(-1,2.5);
\draw[color=blue,densely dashed] (-2,2)--(-2.5,2.5);
\draw[color=blue,thick, ->] (0,0)--(1.5,1.5); 
\draw[color=blue,thick, ->] (0,0)--(-1.5,1.5); 
\draw (0.,-3.1) node {{\small \parbox{8.3cm}{\begin{center}
\emph{Figure 6: What spatial symmetry requires -- beables carry the information about
the B setting to A, and vice versa.}
\end{center}
}}};

\draw (-2.7,-1) node  [rotate=90] {{\footnotesize Time}};
\draw (0.,-0.4) node {M};

\draw (2.5,2.7) node {{\footnotesize $\textup{\textbf{B}}=1$}};
\draw (-2.5,2.7) node {{\footnotesize $\textup{\textbf{A}}=1$}};
\draw (1,2.7) node {{\footnotesize $\textup{\textbf{B}}=0$}};
\draw (-1,2.7) node {{\footnotesize $\textup{\textbf{A}}=0$}};

\draw (2,0.8) node {{\scriptsize\parbox{2cm}{{\scriptsize Polarizing cube\\ set at angle $\beta$}}}};
\draw (-2,0.8) node {{\scriptsize\parbox{2cm}{Polarizing cube\\ set at angle $\alpha$}}};

\end{scope}
\end{tikzpicture}
\end{center}}

\newcommand{\figvi}{\definecolor{orange}{rgb}{1,0.5,0}
\begin{center}
\begin{tikzpicture}
\begin{scope}[scale=1.2]

\draw[color=black,ultra thin] (-3.2,-3.2)--(-3.2,3.2)--(3.2,3.2)--(3.2,-3.2)--(-3.2,-3.2); 
\draw[color=black,ultra thin] (-2.5,-2.5)--(1.5,1.5); 
\draw[color=black,ultra thin] (1.5,-1.5)--(-2.5,2.5); 
\draw[color=black, ->] (-2.5,-1.5)--(-2.5,1.5); 
\draw[color=orange, ultra thick, ->] (1.75,0.25) .. controls (1.75,0.75) .. (1.2,1); 
\draw[color=orange, ultra thick, ->] (1.75,-0.25) .. controls (1.75,-0.75) .. (1.2,-1); 
\draw[color=orange, ultra thick, ->] (-1.5,0.25) .. controls (-1.5,1.4) and (0.25, 1.25) .. (0.9,1.1); 
\draw[color=orange, ultra thick, ->] (-1.5,-0.25) .. controls (-1.5,-1.4) and (0.25, -1.25)
.. (0.9,-1.1);
\draw[color=gray,ultra thick] (2.25,1.75)--(1.75,1.25)--(1.25,1.75)--(1.75,2.25)--(2.25,1.75)--(1.75,1.25);
\draw[color=gray,thick] (1.75,1.4)--(1.75,2.1);
\draw[color=blue,densely dashed] (2,2)--(2.5,2.5);
\draw[color=blue,densely dashed] (1.5,2)--(1.0,2.5);
\draw[color=gray,ultra thick] (2.25,-1.75)--(1.75,-1.25)--(1.25,-1.75)--(1.75,-2.25)--(2.25,-1.75)--(1.75,-1.25);
\draw[color=gray,thick] (1.75,-1.4)--(1.75,-2.1);
\draw[color=blue,thick] (2,-2)--(2.5,-2.5);

\draw[color=gray] (1,-2.5)--(1.5,-2);

\draw[color=blue,thick, ->] (0,0)--(1.5,1.5); 
\draw[color=blue,thick] (1.5,-1.5)--(0,0); 
\draw[color=red,ultra thick] (0,-0.5)--(0,0.5); 
\draw (0.,-3.75) node {{\small \parbox{8cm}{\begin{center}
\emph{Figure 7: What time-symmetry requires -- a second\\ beable carries information about the B setting to C.}
\end{center}
}}};
\draw (1.8,0) node {{\scriptsize C-beable:[$\gamma$]}};
\draw (-1.5,0) node {{\scriptsize B-beable:[$\beta$]}};
\draw (-2.7,0) node  [rotate=90] {{\footnotesize Time}};
\draw (0.,-0.6) node {{\scriptsize Mirror}};

\draw (0.25,1.75) node {{\scriptsize\parbox{2cm}{{\scriptsize Polarizing cube\\ set at angle $\beta$}}}};
\draw (0.25,-1.75) node {{\scriptsize\parbox{2cm}{Polarizing cube\\ set at angle $\gamma$}}};
\draw (1,-2.7) node {{\footnotesize $\textup{\textbf{C}}=0$}}; 
\draw (2.5,-2.7) node {{\footnotesize $\textup{\textbf{C}}=1$}};
\draw (2.5,2.7) node {{\footnotesize $\textup{\textbf{B}}=1$}};
\draw (1,2.7) node {{\footnotesize $\textup{\textbf{B}}=0$}};

\end{scope}
\end{tikzpicture}
\end{center}}

\title{New Slant on the EPR-Bell Experiment}

\author{Peter Evans\thanks{Centre for Time, Department of Philosophy, Main Quad A14, University of Sydney, NSW 2006, Australia; \href{mailto:peter.evans@sydney.edu.au}{peter.evans@sydney.edu.au}.},  Huw Price\thanks{Centre for Time, Department of Philosophy, Main Quad A14, University of Sydney, NSW 2006, Australia; \href{mailto:huw.price@sydney.edu.au}{huw.price@sydney.edu.au}.}, \& K.~B.~Wharton\thanks{Department of Physics and Astronomy, San Jos\'{e} State University, San Jos\'{e}, CA 95192-0106, USA; \href{mailto:wharton@science.sjsu.edu}{wharton@science.sjsu.edu}.}.}

\date{June 20, 2010}

\begin{document}
\maketitle
\thispagestyle{empty}
\begin{abstract}
\noindent The best case for thinking that quantum mechanics is nonlocal rests on Bell's Theorem, and later results of the same kind. However, the correlations characteristic of EPR-Bell (EPRB)  experiments also arise in familiar cases elsewhere in QM, where the two measurements involved are timelike rather than spacelike separated; and in which the correlations are usually assumed to have a local causal explanation, requiring no action-at-a-distance. It is interesting to ask how this is possible, in the light of Bell's Theorem. We investigate this question, and present two options. Either (i) the new cases are nonlocal, too, in which case action-at-a-distance is more widespread in QM than has previously been appreciated (and  does not depend on entanglement, as usually construed); or (ii) the means of avoiding action-at-a-distance in the new cases extends in a natural way to EPRB, removing action-at-a-distance in these cases, too. There is a third option, viz., that the new cases are strongly disanalogous to EPRB. But this option requires an argument, so far missing,  that the physical world breaks the symmetries which otherwise support the analogy. In the absence of such an argument,  the orthodox combination of views -- action-at-a-distance in EPRB, but local causality in its timelike analogue -- is less well established than it is usually assumed to be.
\end{abstract}

\clearpage
\pagenumbering{arabic}

{\small
\ \\
\begin{quotation}\noindent
``Our intuition, going back forever, is that to move, say, a rock, one has to touch that rock, or touch a  stick that touches the rock, or give an order that travels via vibrations through the air to the ear of a  man with a stick that can then push the rock---or some such sequence. This intuition, more generally, is  that things can only directly affect other things that are right next to them. If A affects B without being right next to it, then the effect in question must be indirect---the effect in question must be something  that gets transmitted by means of a chain of events in which each event brings about the next one  directly, in a manner that smoothly spans the distance from A to B. \ldots

We term this intuition `locality.'

Quantum mechanics has upended many an intuition, but none deeper than this one. And this particular  upending carries with it a threat, as yet unresolved, to special relativity---a foundation stone of our 21st-century physics.'' ([\ref{ref:albert}], p.~32.)\vspace{12pt}
\end{quotation}}

\section{Introduction}


The remarks above give vivid expression to a common view about quantum mechanics (QM), viz., that it reveals the existence of some sort of action-at-a-distance (AAD)  in the physical world, in deep tension with our intuition that causation always acts \emph{locally.} While this view of the implications of QM is not universal, it is popular enough  to be termed an orthodoxy -- the \emph{AAD Orthodoxy,} as we shall call it. 
Our aim in this paper is to present a new challenge to the AAD Orthodoxy. We argue that it is significantly less well-grounded than it is widely assumed to be. We aim to show that the case for AAD in QM  -- and with it, the case for thinking that QM carries ``a threat, as yet unresolved, to special relativity'' -- has not yet been made.  

In the interests of clarity, we wish to distinguish our argument from other challenges to the AAD Orthodoxy. For this reason, we set aside from the beginning the Everett or ``Many Worlds'' interpretation of QM. The status of the AAD Orthodoxy in the Everett picture remains controversial -- see, e.g., [\ref{ref:blaylock}] and [\ref{ref:maudlin}] for recent expressions of conflicting viewpoints on this matter. We take no stand on this issue, and restrict our discussion to single-outcome views of QM. We note, however, that if supporters of the  Everett interpretation subscribe to the AAD Orthodoxy in the conditional sense that they take rival views of QM to be committed to AAD, then they, too, are amongst our targets. 

In a similar spirit, we set aside other objections to the  AAD Orthodoxy, such as the view that the novel spacelike correlations revealed by QM are not  \emph{causal} in nature, or do not involve \emph{action} at a distance; and the view that QM has an epistemic or instrumental function, and hence is not in the business of describing the causal structure of reality.\footnote{See, e.g., [\ref{ref:shim}] and  [\ref{ref:cohen}] for the former viewpoint, and [\ref{ref:fuchs}] for an  introduction  to the latter.} Again, our arguments may be of interest to proponents of such views in various indirect ways; but we shall not ourselves engage with these  rivals to the AAD Orthodoxy.\footnote{Some readers may feel that, rightly or wrongly, the view that QM reveals AAD  is \emph{not} sufficiently widespread to be called an orthodoxy.  But since we are criticising the view, rather than trying to appeal to it as established fact, this is merely a terminological point. Readers should feel free to substitute their own label. We recommend [\ref{ref:norsen}] for a clear recent presentation of the view, as we take it to be.}

\subsection{Background}

The challenge that QM poses to local causality, and hence to special relativity, stems from a feature of the theory called \emph{entanglement.} The relevant consequences of entanglement  emerge in the so-called Einstein-Podolsky-Rosen (EPR) experiment, and especially in John Bell's [\ref{ref:bell}] famous analysis of a version of EPR proposed by David Bohm. We refer to these cases, generically, as EPRB experiments. (We describe a particular example in \S 2 below.)

EPRB experiments display distinctive patterns of correlations between the outcomes of measurements on separated pairs of quantum systems. These correlations seem at first sight to invite explanation in terms of a ``common cause'' -- i.e., in terms of some correlation between the underlying properties of the two systems, established when they interact in their common past, and responsible for the outcomes of the measurements concerned. However, Bell showed that under seemingly uncontroversial assumptions, such a common cause explanation is not possible, in some cases. In these cases, a common cause explanation would require that the joint probabilities of measurement outcomes satisfy a condition  -- Bell's Inequality --  whose violation  is predicted by QM (and now well confirmed by experiment).

Bell's result is often regarded simply as a difficulty for the hidden variable (HV) program in QM, but as Albert and Galchen emphasise in the piece from which we quoted above, to see it this way is greatly to underestimate its significance. Bell's analysis shows that in many cases, the joint probability of pairs of outcomes for possible measurements on each of two entangled particles, cannot be expressed as a product of a probability for each outcome individually, if the latter probabilities are not allowed to depend in each case on the setting chosen for a measurement on the other particle.\footnote{While the probabilities here are conditional on any hidden variables which may be present in the systems in question, the result does not \emph{require} that there be any such variables.} 
 This interdependence in the joint probabilities is  the main basis for the view that QM entails AAD. (We refer readers to Bell's late paper [\ref{ref:bell2}] and to [\ref{ref:norsen}], for clear accounts of these points.)

\subsection{Outline of the argument}

Our challenge to the AAD Orthodoxy goes like this. We begin with an EPRB experiment involving polarization measurements on photon pairs. We then note that the characteristic pattern of correlations between measurement settings and outcomes in this experiment also arises in a familiar experiment involving repeated polarization measurements on a single photon. In this case, which we call the \emph{Sideways} EPRB experiment (SEPRB), the two polarization measurements have a \emph{timelike} separation, rather than the \emph{spacelike} separation possible in EPRB. The correlations in SEPRB are not normally thought be a manifestation  of AAD -- on the contrary, they are usually assumed to have a ``local'' causal explanation, given by the standard QM description. But if SEPRB manages to avoid AAD, despite exhibiting the same pattern of correlations as EPRB, it is interesting to ask how it does so.

Investigating this question, we present the following three options:
\renewcommand{\labelenumi}{\Roman{enumi}.}
\begin{enumerate}
\item SEPRB  involves a kind of AAD, too, in which case AAD is much more widespread in QM than has previously been appreciated (and  does not depend on entanglement, as usually construed). 
\item The means of avoiding AAD in SEPRB extends in a natural way to  EPRB, eliminating AAD  (and hence its conflict with special relativity) in these cases, also.
\item SEPRB is more disanalogous to EPRB  than is evident from the mathematics, so that
the means of avoiding AAD in the former does not extend to the latter.
\end{enumerate}
Option II amounts to a direct challenge to the AAD Orthodoxy. From the point of view of the Orthodoxy, then, the acceptable options are I and III.

But neither of these options (I or III) is ``cost-free''. On the one hand, Option I conflicts not only with the standard view of where AAD originates in QM, but also -- as the other side of the same coin -- with the widely accepted ``intuitive'' picture of the explanation of quantum correlations in the SEPRB cases (and many others like them). On the other hand, Option III requires rejection of some attractive symmetries, which otherwise support the analogy between SEPRB and EPRB. So the AAD Orthodoxy has some work to do,  to convince us that one of these choices is really more plausible than Option II.

We emphasize that this argument does not show that the AAD Orthodoxy is \emph{false.} But it does show that it cannot be regarded as \emph{well-justified,} in the absence of a case -- so far largely non-existent -- for preferring Options I or III to Option II. The Orthodoxy may perhaps be \emph{true,} in other words; but it is not presently \emph{well-grounded.}

As we note below, symmetry-based objections to the AAD Orthodoxy have been offered before. Our argument is novel in two main respects. First, it exploits a timelike variant of the kind of experiment (i.e., EPRB) usually supposed to provide the strongest case for AAD in QM. This enables our symmetry argument to take an unusually direct form. In effect, we simply point out that whatever ontology underlies the  local causality  in SEPRB will do the same job in EPRB, so long as the symmetries hold. Secondly, and relatedly, the EPRB/SEPRB comparison enables us to appeal to a second kind of symmetry argument, in addition to the \emph{time-}symmetry central to earlier proposals.  

Finally, we emphasize that we do not claim that these symmetry arguments cannot be challenged. On the contrary, we shall stress at several points that one option for the AAD Orthodoxy, in the face of our argument, is to reject these symmetries. But a would-be defender of the Orthodoxy who announces that she takes this option (i.e., Option III, in the list above) has not given us an argument, but only an IOU. To defend the Orthodoxy (by this route), she needs to \emph{make a case} that the symmetries fail. In the absence of such an argument, the AAD view is no more than one hypothesis among several live options.

\section{The experiments}


\subsection{Standard EPRB}\vspace{12pt}

\figi

Figure 1 shows a familiar version of an EPRB experiment, performed with pairs of photons. Photons emitted in a back-to-back geometry from a two-photon-decay at M (say, a $2s-1s$ transition in Hydrogen) reach ideal polarizing cubes {A} and {B} whose orientations $\alpha$ and $\beta$, respectively, may be freely and independently chosen by two experimenters.   Subsequent detection of each photon has two possible results; transmission (\textbf{A}=1, \textbf{B}=1) or reflection (\textbf{A}=0, \textbf{B}=0).   (Bold capitals represent Boolean variables.) Before measurement, the photons are described by an entangled state of zero total angular momentum, usually interpreted to imply that neither photon has a determinate polarization. After one measurement is made, however, the typical interpretation is that both photons acquire a determinate polarization based upon the orientation of the polarizer and the measurement result. The joint probability $Pr(\textup{\textbf{A}}=\textup{\textbf{B}})$ of the same result at {A} and {B} is known to be $\cos^{2}(\alpha - \beta)$, violating Bell's Inequality, and therefore impossible to replicate with hidden variables at M that do not depend on the settings $\alpha$ and $\beta$.


\subsection{Sideways EPRB}

Figure 2 shows the spacetime diagram for  the Sideways EPRB (SEPRB) experiment.  The latter polarizing cube at B is exactly the same as in EPRB; it is set at an angle $\beta$%
, and the measurement \textbf{B} indicates whether the photon was transmitted or reflected in the cube.\vspace{12pt}

\figii

For any B-measurement to take place, a photon must first reflect from the mirror.  In order to reach the mirror, the photon must pass through an earlier polarizer at C.  This polarizer is set at an angle $\gamma$ (meaning that transmitted photons have a polarization aligned with $\gamma$).  There are two ways for an experimenter to introduce a single photon along the appropriate trajectory: (1) it can be injected from the right (\textbf{C}=1), followed by a successful transmission through the polarizing cube (giving it a polarization aligned with $\gamma$), or (2) it can be injected from the left (\textbf{C}=0), followed by a successful reflection in the polarizing cube.  The latter case leaves the photon with a polarization aligned with $\gamma+\pi/2$.  Here \textbf{C} is a boolean variable that does not represent an experimental outcome, but rather an input choice made by the experimentalist.

If the experimenter chooses \textbf{C}=1, it is well known that the probability of \textbf{B}=1 is given by \cscb.  On the other hand, if the experimenter chooses \textbf{C}=0, then the probability of \textbf{B}=0 is now \cscb.  So the correlation between the experimenter's choice \textbf{C} and the measurement outcome \textbf{B} is $Pr(\textup{\textbf{B}}=\textup{\textbf{C}}) = \cos^{2}(\beta - \gamma)$.  The polarization of the input to the first polarizing cube is irrelevant, so long as one considers a null result at B to be no experiment at all.  Note that the mirror is included only for heuristic purposes, to emphasise the similarities between this case and the standard EPRB experiment in Figure 1. We could equally place the first polarizing cube at T, and remove the mirror.

\subsection{Comparing the experiments}
SEPRB involves exactly the same pattern of correlations between distant measurement events as EPRB. In both cases, the widely-separated events have a correlation of the form \csab , where $\alpha$ and $\beta$ are  freely and independently variable by the  experimenters concerned. As is well known (and experimentally well confirmed), the correlations in EPRB violate Bell's Inequality, and it is this fact, in combination with the assumptions of Bell's derivation of his Inequality, that is taken to entail that EPRB involves AAD.

Since SEPRB displays the same correlations as EPRB, it, too, must violate Bell's Inequality. Does this imply that SEPRB involves AAD? No, or at least not automatically, for it may well be that one or more of the assumptions required for the derivation of Bell's Inequality in EPRB do not hold in SEPRB.  This is the possibility we wish to investigate, for we are interested in the question as to whether the way in which these assumptions are thought to fail in SEPRB might throw light on how they might fail in EPRB, too (thus blocking the argument to AAD in that case). To clarify our argument at this point, we now introduce a brief description of the structure of the derivation of Bell's Inequality.


The basis of the derivation of Bell's Inequality is depicted in Figure 3. 
There are two crucial assumptions.  The first, \textit{Independence,} requires that  any HVs $\lambda$ be independent of the choice of measurement settings $m_2$ and $m_1$ at $\textup{M}_{2}$ and $\textup{M}_{1}$, respectively. The second, \emph{Locality,} specifies that any correlation between $m_2$ and the measurement outcome at  $\textup{M}_{1}$ be ``screened off'' by the value $\lambda_{R}$ of $\lambda$ in some spacetime region R, disjoint from the past light cone of $\textup{M}_{2}$. In other words, the measurement outcome at  $\textup{M}_{1}$ is probabilistically independent of $m_2$, when we hold fixed the value of $\lambda_{R}$.\footnote{An analogous condition is also assumed in the opposite direction, applying to correlations between the setting $m_1$ and the measurement outcome at  $\textup{M}_{2}$. And both conditions normally refer also to ``unhidden'' variables relevant to the outcomes of $\textup{M}_{1}$ and $\textup{M}_{2}$ -- indeed, it is crucial that the screening-off condition hold fixed \emph{everything} in R, and not merely HVs. For simplicity we have here suppressed these points, but again refer readers to  [\ref{ref:bell2}] and [\ref{ref:norsen}] for details. (Our Figure 3 is based on one given by Bell in [\ref{ref:bell2}].)}
(We may think of the measurements $\textup{M}_{2}$ and $\textup{M}_{1}$ as corresponding to those at A and B, respectively,  in EPRB.)\vspace{12pt}

\begin{center}
\begin{tikzpicture}
\begin{scope}[scale=1.2]

\draw[color=black,ultra thin] (-3.2,-1.5)--(-3.2,2)--(3.2,2)--(3.2,-1.5)--(-3.2,-1.5); 
\draw[color=black,thin] (-1,-1)--(1.5,1.5)--(3,0); 
\draw[color=black,thin] (1,-1)--(-1.5,1.5)--(-3,0); 

\draw[color=black,ultra thin,densely dashed] (-2,1)--(-1,1); 
\draw[color=black,ultra thin,densely dashed] (-2.5,0.5)--(-0.5,0.5); 

\draw (1.55,1.7) node {{\footnotesize $\textup{M}_{2},m_2$}};
\draw (-1.45,1.7) node {{\footnotesize $\textup{M}_{1},m_1$}};
\draw (-1.95,-0.1) node {{\footnotesize $\textup{M}_{3},m_3$}};
\draw (-1.98,-0.37) node {{\Huge .}};

\draw (-1.5,0.75) node {{\footnotesize R\hspace{10pt}$\lambda_{R}$}};

\draw (0,-0.4) node {{\footnotesize $\lambda$}};
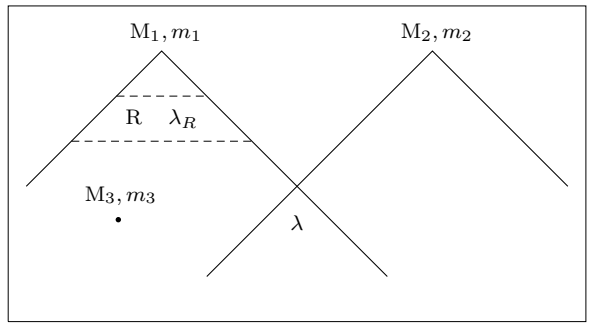
\draw (0.,-2) node {{\small\emph{Figure 3: Independence and Locality in Bell's Theorem}}};

\end{scope}
\end{tikzpicture}
\end{center}

It is important to note that although the fact that $\textup{M}_{2}$ and $\textup{M}_{1}$ have spacelike separation plays a major part in the intuitive justification of Independence and (especially) Locality, it plays no role whatsoever in the derivation of Bell's Inequality, once these assumptions are in place. Hence, if there were a measurement $\textup{M}_{3}$ with setting $m_{3}$ in the past light cone of $\textup{M}_{1}$ such that $\lambda_{R}$ was independent of the choice of $m_{3}$ (Independence) and any correlation between $m_3$ and the outcome at  $\textup{M}_{1}$ was screened off by
$\lambda_{R}$ (Locality), then Bell's Inequality could be derived for this case, too. So if the actual correlations in such a case turned out to violate Bell's Inequality, we would have a demonstration that at least one of Independence or Locality must fail. If Independence was assumed to hold, then it would have to be Locality that failed. This would entail a form of AAD -- in this case, action-at-a-\emph{timelike}-distance.\footnote{Intuitively, action at a \emph{spatial} distance allows a change at one place to produce a change at a remote place, with no change at all on paths in between. Action at a \emph{temporal} distance allows a change at one time to produce a change at a remote time, with no change at times in between. Both conflict with the intuition that causation acts ``locally'' and ``continuously''.}  In other words, it would involve a direct influence of the choice of $m_3$ on the outcome at  $\textup{M}_{1}$, bypassing conditions in R, of the kind prohibited by Locality.
 
Applying this hypothetical form of reasoning to SEPRB, with C in place of $\textup{M}_{3}$ and B in place of $\textup{M}_{1}$, 
it entails  that if Independence were to hold in SEPRB, we could use the fact that the SEPRB correlations (being the same as the EPRB correlations) do not satisfy Bell's Inequality to infer that there is AAD in SEPRB. It would then have turned out that AAD in QM does not require
a second particle, let alone entanglement as normally conceived.

Few people would  interpret SEPRB in this way, of course.\footnote{Though see [\ref{ref:brukner}] for an apparent exception.} On the contrary,  we find it natural to assume that in SEPRB the state of the photon between the polarizers \emph{does} depend on the setting $\gamma$, so that Independence fails. This is why, despite some formal similarities between SEPRB and EPRB, the correlations of the former seem unproblematic, while the correlations of the latter seem ``spooky'' -- why the latter but not the former seem to require AAD.

\subsection{The need for beables}

At this point  it is crucial to observe that this way of avoiding AAD in SEPRB requires that we regard the photon as possessing a ``beable'', in Bell's terminology, to provide the intermediary between the setting at {C} and the measurement at {B}. Whether the quantum state itself can play this role depends on how it is interpreted. A purely epistemic state is not a beable, for example. Nor is a state which merely describes an ensemble of systems, leaving it open whether the individual photon possesses a property sufficient to carry the information about  {C}'s setting from {C} to {B}.

In order to avoid the conclusion that SEPRB involves AAD, we therefore need to make explicit that there is a beable \textit{of some kind} -- perhaps the quantum state, appropriately interpreted, perhaps some other ontic state -- capable of carrying the information about a measurement setting from {C} to {B}. Since it does the job of carrying information from {C}, we shall call it the \textit{C-beable.} The notation ``C-beable:[$\gamma$]'' in Figure 4 thus represents the fact that the C-beable carries the information that {C}'s polarizer setting is $\gamma$. We shall say that such a beable provides a \emph{Locality Model} for SEPRB.\footnote{For completeness, we note that in principle one might devise a Locality Model in which the information about  the measurement setting at C is carried to B by something other than a beable \emph{of the photon;} i.e., with a different mechanism  altogether for ``action-by-contact'' between C and B. This option might  evade our symmetry arguments below, but we think it unlikely to appeal to any defender of ``intuitive causality'' in SEPRB.}\vspace{12pt}

\figiii

The next step is to appeal to symmetry considerations, to call attention to the possibility that a Locality Model for SEPRB might transform into  a Locality Model for EPRB, too, under application of some suitable symmetry. Our goal here is to show that \emph{if} one accepts the following two assumptions -- (i) that there is a Locality Model for SEPRB, and (ii) that the beables of this Locality Model respect certain symmetries -- \emph{then} one must accept that there is a Locality Model for EPRB, too. (In other words, one must accept that there are beables to mediate what would otherwise be AAD in SEPRB, in conflict with the AAD Orthodoxy.) This leaves the AAD Orthodoxy with the choice of rejecting assumption (i) or rejecting assumption (ii); and our challenge to the Orthodoxy is to say \emph{which} assumption it rejects, and \emph{why.} 

 It turns out that there are two  largely independent symmetries, either one of which could play the part required in this argument. We present them  in turn.


\section{The symmetry considerations}

\subsection{The action symmetry}

If beables can explain the correlations seen in SEPRB, it is natural to ask whether the identical correlations in EPRB might be explained in an analogous manner. Indeed, these two experiments are more than superficially similar; they span bounded regions of spacetime with precisely the same electromagnetic action $S$. It should therefore be no surprise that the experimental correlations in EPRB and SEPRB are identical, as our most advanced theory of these interactions -- quantum electrodynamics (QED) -- reduces the joint probability to a functional integral of the classical action:
\begin{equation}
\label{eq:prob}
P[A(t_0),A(t_f)]=\left|\int {\cal D}A e^{iS[A]/\hbar}\right|^2.
\end{equation}

Here $A$ is the electromagnetic 4-potential, and the integral is over all field configurations consistent with the initial boundary $A(t_0)$, the final boundary $A(t_f)$, and any spatial boundaries (such as the mirror). If we post-select the EPRB measurement outcome to match the input choice in SEPRB (\textbf{A}=\textbf{C}), and also filter the photons in EPRB such that they are the same wavelength, then there is an exact ``action symmetry'' (S-symmetry) between these two experiments for any given outcome \textbf{B}. (This is related to ``crossing symmetry'' in quantum field theory.)

To see this explicitly, consider the following action-preserving permutation of SEPRB.  Taking the ``bottom half'' of the rectangular spacetime region bounded by the mirror and the polarizers (the shaded region in Figure 4), one can flip the contents of this sub-region from left-to-right, and then again  future-to-past, without changing the global action.  Then, one can imagine removing this sub-region from its location \textit{before} the unshaded ``top half'' of the experiment, and re-affixing it to a new spacetime location \textit{to the left of} the ``top half'', again without changing the action.  This procedure results in the EPRB geometry, where {C} $\to$ {A}.  The only complication is at the vertex where the photon worldlines meet, but notice that both the mirror (in SEPRB) and the two photon decay (in EPRB) serve to correlate the polarization of the two photons in the same manner, so no action boundary terms are introduced or lost in this process.\footnote{For ease of exposition, we are now treating the reflected photon in SEPRB as a ``new'' photon, distinct from the incident photon, as reflection in QED is typically interpreted.}

Given that $S$ is the same, the only difference between the use of Eqn.~(\ref{eq:prob}) in the two experiments (for a given \textbf{B} outcome) is whether the \textbf{C} input condition is constrained by $A(t_0)$ (SEPRB) or the \textbf{A} output condition is constrained by $A(t_f)$ (EPRB). But as far as Eqn.~(\ref{eq:prob}) is concerned, this is no difference at all; joint probabilities don't distinguish past from future.  Eqn.~(\ref{eq:prob}) thus offers a natural explanation of the fact that both experiments display the same correlations; they result from the same QED mathematics.\footnote{The use of these joint probabilities in QED may not be familiar to 
readers used to the conditional probabilities of non-relativistic
QM, but Leifer [\ref{ref:leif06}] has demonstrated an isomorphism between these two types of
probability, and arrives at a
similar conclusion concerning a symmetry between EPRB-like and SEPRB-like
experiments.} (Note that the addition of any physical faster-than-light communication channel between A and B in EPRB would break the S-symmetry.)

Does this symmetry in the mathematics \emph{entail} a corresponding symmetry in the ontology underlying EPRB and SEPRB? That conclusion would be too strong, in our view. For one thing, QED does not provide an accepted physical interpretation of these correlations, so we should be more than usually cautious about reading ontology from, or into, the mathematics, in this case. For another, we want to leave open the option that there simply is no underlying ontology, in either case. 

A weaker conclusion suffices for our purposes, however. The crucial point is that the symmetry in the mathematics offers the prospect of a clear road to anyone who wishes to exploit the S-symmetry to transform a Locality Model for SEPRB into a Locality Model for EPRB -- at least, a clear road so far as the mathematics is concerned. So long as the ontology of a Locality Model for SEPRB lives in spacetime, this action-preserving permutation of the \emph{geometry} (from that of SEPRB to that of EPRB) provides a simple QED-inspired template for a corresponding permutation of the \emph{ontology.} While QED provides no \emph{guarantee} that the ontology of the real world respects this symmetry, it certainly raises the possibility that it might do so. (One might even say -- not wishing to be too bold -- that it offers us a \emph{hint} that it might do so.) This is the possibility that the AAD Orthodoxy must find reason to reject, if it is to make a case for the claim that there is a Locality Model for SEPRB but \emph{not} for EPRB.\footnote{One of the tasks that falls on the Orthodoxy, if it rejects S-symmetry at the level of \emph{ontology,} is to explain the evident success of QED's S-symmetric \emph{formalism.}} 

Our challenge to the AAD Orthodoxy is thus to propose some such roadblock -- some reason for thinking that the ontology of SEPRB and EPRB does not reflect the symmetry of the mathematics.
 In the absence of such a roadblock, the transformation  of a Locality Model for SEPRB into a Locality Model for EPRB proceeds as follows. It begins with the intuitive interpretation of SEPRB, as in Figure 4, and then applies the S-symmetry permutation described above, to produce Figure 5.

In Figure 5, the A-beable is correlated with a future measurement setting. This is certainly counterintuitive, in the sense that it conflicts with what physicists often refer to as ``intuitive causality''. While it has often been noted that such a beable would allow us to violate Bell's Inequality
with local hidden variables (see, e.g., [\ref{ref:argaman}],  [\ref{ref:costa}]--[\ref{ref:cram86}],  [\ref{ref:hokk}], [\ref{ref:mill96}],  [\ref{ref:mill97}], [\ref{ref:price84}]--[\ref{ref:price08}], [\ref{ref:reit}], [\ref{ref:sutherland83}]--[\ref{ref:sutherland}], [\ref{ref:wharton07}], [\ref{ref:wharton09}]), most commentators have thought the cost too high to pay. We return below to the question of what this cost actually amounts to.\vspace{12pt}

\figiv

For the moment, ignoring such issues, we focus simply on the evident symmetry between EPRB and SEPRB. Our point is that rejecting A-beables in EPRB, while accepting C-beables in SEPRB, requires one to reject at the ontological level the S-symmetry revealed at the mathematical level by the identical electromagnetic action of both EPRB and SEPRB.\footnote{It seems likely that a rejection of S-symmetry would also require non-trivial modifications to QED itself.}

Still, another symmetry remains violated in Figure 5; a spatial symmetry. To enforce a strict parallel between Figures 4 and 5, we must post-condition on  outcome \textbf{A}, leaving outcome \textbf{B} to be determined by the joint probability Pr(\textbf{A}=\textbf{B}). But such one-sided post-conditioning should not be sufficient to break the spatial symmetry between {A} and {B}. In order to recover spatial symmetry, one needs to treat the future settings $\alpha$ and $\beta$ on an equal footing, leading to both A-beables and B-beables, as shown in Figure 6 (instead of Figure 5).
\figv
\vspace{6pt}

\figvi

Under S-symmetry, recovering a strict spatial symmetry in EPRB is equivalent to recovering a strict temporal symmetry in SEPRB. This is easily seen by applying S-symmetry to Figure 5 (the {A}$\leftrightarrow${C} map) yielding the time-symmetric Figure 7. Figures 6 and 7 represent ontologies that each require beables sensitive to both measurement settings.  The spatially-symmetric geometry of EPRB results in a fully spatially-symmetric ontology; the time-symmetric geometry of SEPRB results in a fully time-symmetric ontology; the two ontologies are also S-symmetric with respect to each other.\

S-symmetry thus provides one path by which a Locality Model for SEPRB might be transformed into a Locality Model for EPRB. We stress, once again, that our principle conclusion at this point  is not that the path is indubitably open, but rather that onus lies with the AAD Orthodoxy to show that it is closed. 

\subsection{Time-symmetry in SEPRB}
In the proposal above, time-symmetry in a Locality Model for SEPRB emerged from the application of two other symmetries: S-symmetry, and spatial symmetry within EPRB. Our second symmetry proposal reverses the order of priority. It begins, like the first, by assuming a Locality Model for SEPRB; but then appeals directly to time-symmetry, noting that unless the Locality Model in question is time-asymmetric, it also yields a Locality Model for EPRB. We stress in advance that this second proposal does not depend on -- or imply -- strict S-symmetry, and hence is impervious to challenges based on objections to S-symmetry.\footnote{
A weaker claim is true in reverse. Since S-symmetry maps a spatial-symmetry to a time-symmetry, a principled objection to the time-symmetry in question (without a corresponding reason to doubt the spatial-symmetry) would be an objection to S-symmetry, too.}

We noted above that the intuitive strategy for avoiding AAD in the SEPRB experiment requires a C-beable to encode the information about the setting of polarizer C, making this information available at B. We now raise the question: Is there also a B-beable which encodes the information about the setting of polarizer B (as in Figure 7)? If the usual quantum state $\psi$ is regarded as a beable, and we assume the projection postulate, then the answer to this question is ``No''. As many writers have noted, this means that such interpretations are time-asymmetric.  Some have seen this as an objection, others as a reason for thinking that physics is not time-symmetric, at a fundamental level. But as has also been noted, time-symmetry can be restored by introducing a second wave function considered as carrying information about future interactions [\ref{ref:aharonov}, \ref{ref:sutherland}, \ref{ref:wharton07}] or by using second-order (in time) wave equations constrained by both past and future interactions [\ref{ref:wharton09}].  In these cases, provided that the wave functions have the status of beables, we do have B-beables.

Again, our interest is not in the usual quantum description as such, but in the beables -- whatever they may be --  required to avoid AAD, in the SEPRB experiment. Our point is simply that unless a Locality Model for SEPRB is time-asymmetric, it must contain B-beables, as well as C-beables.

Let us suppose for the moment that our Locality Model for SEPRB is time-symmetric in the manner shown in Figure 7. It is now a simple matter to show that this use of beables correlated with future settings also provides the resources to remove AAD from EPRB.  As shown in Figure 6, beables correlated with future measurement settings are available to the source, M.  Then, provided we require that the source produce pairs of particles with correlated beables, that information will be available to both future measurements at {A} and {B}. Thus, as we wanted to show, a time-symmetric Locality Model for SEPRB easily extends to a Locality Model for EPRB.

\section{The basic trilemma}

We have shown that there are two symmetry-based strategies for extending a Locality Model for SEPRB to a Locality Model for EPRB. The first strategy appeals to S-symmetry directly. It then invokes spatial symmetry in EPRB, in order to produce a spatially-symmetric Locality Model for EPRB, and thence (by a second application of S-symmetry), a time-symmetric Locality Model for SEPRB. The second strategy appeals to time-symmetry in the context of the original Locality Model for SEPRB. It points out that time-symmetry in this context requires  B-beables, and  that these may then be used to generate a Locality Model for EPRB, too.

With these strategies  on the table, we are confronted immediately by the three options we identified in \S1.2, and which we shall now call the \emph{Basic Trilemma:}

\renewcommand{\labelenumi}{\Roman{enumi}.}
\begin{enumerate}
\item \textbf{More AAD than we thought.}  We can reject the project of finding a Locality Model for SEPRB altogether, and conclude that AAD  is much more widespread in QM than is usually supposed. We thus reject the intuitive causal model for SEPRB (and presumably a vast number of similar cases).

\item \textbf{Less AAD  than we thought.}  We can put our faith in the project of exploring symmetric Locality Models, which account in an action-by-contact (and hence potentially Lorentz-invariant) way for both the EPRB and SEPRB cases.

\item \textbf{As much AAD  as we thought, but less symmetry.}  We can continue to maintain that there is  local causality in SEPRB but AAD in EPRB, provided we  reject \emph{both} the symmetries that otherwise take us from a Locality Model for SEPRB to a Locality Model for EPRB. 
\end{enumerate}
We take Option III to be the standard and ``intuitive'' view, at least among proponents of the AAD Orthodoxy (whose intuitions have long since become hardened to AAD). We have not excluded this option, but we have argued that it incurs a cost, in symmetry terms. If we choose it, we commit ourselves to the view that the ontology that underlies local causality in SEPRB is neither \emph{S-symmetric,} in the sense embodied in  the formalism of QED, nor \emph{time-symmetric.}\footnote{Though see also \S6 on the latter point.} 

At this point, proponents of the AAD Orthodoxy thus have a choice to make. Should they renounce Option III in favour of Option I, or should they make a stand on the failure of these symmetries? For the purposes of this paper, we shall not pursue those who make the former choice (i.e., Option I). They move in the direction of a less realist conception of QM, and hence off our map, for present purposes. 

Our main conclusion is that those who remain -- in other words, those advocates of the AAD Orthodoxy who reject Option I -- now owe us a defence of Option III. That is, they need an argument for preferring Option III to Option II: a reason for thinking that the ontology underlying the true Locality Model for SEPRB \emph{does not} respect the symmetries that would transform it into a Locality Model for EPRB.

\subsection{An intuitive defence of Option III?}


It might seem that one way to produce such an argument would be simply to appeal to ordinary causal intuitions, something like this:\footnote{We borrow some of the following text from an anonymous referee.}
\begin{quotation}
\noindent ``We look at the sideways EPRB case and say: the intuitive explanation involves a photon propagating from the first measurement to the second, where the state of the photon depends on the setting and outcome of the first measurement. We look at the normal EPRB case and say: the intuitive explanation involves two photons, propagating in opposite directions to their respective measuring apparatuses and carrying no information about the settings of the apparatuses. The first intuition provides our reason for rejecting Option I. The second intuition then provides our reason for favouring Option III; for these two intuitions are only compatible if the symmetries are not a good guide to the relation between the underlying ontology in the two cases.'' 
\end{quotation}

However, in relying on causal intuitions in support of Option III, this argument is limited in a crucial respect. The AAD Orthodoxy has already accepted that QM requires us to jettison another of our long-cherished causal intuitions, viz., the belief that there is no action-at-a-distance. This makes it difficult, to say the least, for the Orthodoxy to take the high ground in appealing to \emph{other} causal intuitions.  For the obvious rejoinder will be, ``Why not sacrifice one of \emph{those} sacred cows, rather than the principle that there is no AAD?'' 

Both sides in this debate agree that QM shows that ``something has to give'': some element in our intuitive causal picture must go. The AAD Orthodoxy is committed to a particular view about what should go, i.e., that we must sacrifice local causality. But if this is to be a reasoned position, rather than a mere expression of taste, it needs to be based on an argument for Option III that breaks out of the circle of appeals to intuition. 

The same is true for the AAD Orthodoxy's opponents, of course. But they have an obvious candidate for an independent arbiter, in the tension between AAD and special relativity. An appeal to special relativity in support of Option II is an appeal to physics, not merely to ordinary intuitions about causality.

So intuition alone cannot provide an adequate defence of Option III. This does not imply, of course, that no other defence can be found.  (We mention some other candidates in \S7 below.) Our point is  simply that such an argument needs to be found, if the AAD Orthodoxy is to be regarded as well established (and if Option I is rejected). Without such an argument, the case for the Orthodoxy is seriously incomplete.

\section{Avoiding the trilemma?}

Some readers may feel that the apparent need to make a choice between these three options only arises if we have not properly learnt the lessons of QM. We anticipate a range of objections of this kind. From one side, it may be claimed that our whole discussion is premissed on excessive (``na\"{\i}ve'') realism about the quantum world: ``Avoid such realism, as QM shows us that we must, and your trilemma will melt away.'' From a different angle, it may be objected that we have not understood that the only \textit{real} ontology in the quantum world is the wave function, as recommended by the Everett view; and that this view simply transcends the concerns about locality on which our discussion has been based.

We cannot do justice here to all such objections, and in any case stipulated at the beginning that we were setting these views to one side, for the purposes of this paper. However, we would like to stress two general points. First, since such objections (of the former kind, especially) align themselves most naturally with Option I, we emphasise the strength of ordinary causal intuitions, in the SEPRB case. One way to stress this point is to note that SEPRB can easily be used for signalling, and hence for remote control of macroscopic processes.\footnote{The crucial disanalogy with EPRB, which prevents signalling in the latter case, is that in SEPRB we have direct control of the input at {C}; whereas in EPRB we cannot control the outcome at {A}, fixing it only by post-selection.} Rejecting the intuitive causal picture in such cases, whether on antirealist, Everettian or other grounds, is not a step to be taken lightly.\footnote{It seems to commit us to the view that an effect may be separated from its cause by an arbitrarily long period of time in which the world is \emph{exactly} as it would be, had the cause  not occurred -- for this is what failure of Locality amounts to, in the timelike case, as Figure 3 illustrates.} Our central argument has been that if it is not rejected in these cases, it need not (and should not) be rejected in EPRB either; unless some reason can be found for breaking  some otherwise plausible symmetries.

Second, we note that the relative merits of the views of QM which lie behind these objections -- e.g., of the Everett view on one side, or of instrumentalist views on the other -- depend on the disadvantages of alternative approaches. In particular, they depend on the viability or otherwise of HV approaches. Our claim is that in so far as the apparent disadvantages of HV views rest on Bell's Theorem (and, indeed, on other No Hidden Variable results relying on similar assumptions about the independence of HVs from future measurement settings), the case for the negative may have been seriously overstated. It has not usually been felt that cases such as SEPRB provide any intuitive difficulties for a HV explanation; and yet these cases, combined with the symmetries, seem to provide the resources needed to account for EPRB cases as well. Unless some argument can be found why the symmetries \emph{need} to be broken, this aspect of any case against HV views is considerably weaker than it is usually assumed to be.

\section{The classical objection}

In partial defense of the acceptability of Option III in our trilemma, it might be claimed that the time-asymmetry of a Locality Model for SEPRB with C-beables but no B-beables need not be fundamental; so that such a model need not conflict with the assumed time-symmetry of the fundamental ontology. Justification for such a claim might be sought in
classical electromagnetism (CEM), which uses manifestly time-symmetric equations to yield the same \cscb\ ratio between the output intensity at \textbf{B}=\textbf{C} and the intensity after the polarizer at C. This is Malus's Law, and such a correlation seems to have no need for a fundamental time-asymmetry (or retrocausality, for that matter).

But accomplishing this feat for CEM is not the same as explaining single-photon correlations.  Mapping the continuous parameter $(\beta - \gamma)$ onto a continuous set of possible outcomes in CEM is not analogous to mapping the same parameter onto the two actual outcomes \textbf{B}=0 or \textbf{B}=1 in the single-photon case.\footnote{This point can be strengthened if we accept an argument proposed in 
[\ref{ref:price10}] 
for the conclusion that the discreteness of the quantum description introduces a new kind of boundary-condition-independent correlation between beables and settings; and hence guarantees retrocausality, in a Locality Model for SEPRB, if the ontology (i) is time-symmetric, and (ii) does not restore continuity in the possible outputs, e.g., by means of an ontic  wave function.}  One could accomplish this continuous $\to$ discrete transition with something like QM's projection postulate, but this would break the time-symmetry of the ontology itself.

It is true that we cannot entirely rule out the possibility that someone might suggest a new level of description, where the time-asymmetric use of C-beables but not B-beables would merely be a product of special initial conditions (and not a time-asymmetric ontology). To put this possibility in perspective, however, we offer the following comment.

At familiar scales, physics is dominated by special initial conditions (now widely believed to be associated with the low entropy condition of the universe, after the big bang). The resulting asymmetries are statistical in nature, and associated with mean behaviour of  very large numbers of interactions. There are two possible views about what we should expect when we study individual interactions themselves. One is that the familiar initial-condition-dependent asymmetries will be absent at this level, the other is that they will emerge again, in some new statistical level underlying our current description. It is important to recognise that both paths remain open to physics, but even more important not to confuse them.

In particular, it is important that when we are exploring the first path -- investigating quantum phenomena at what we take to be the most fundamental possible level of description -- we not let our thinking be guided by assumptions that belong only to the second path. So long as we stay on the first path, in other words, we are not entitled to attribute the time-asymmetric use of C-beables but not B-beables to special initial conditions. It cannot be treated as a ``harmless'', non-fundamental time-asymmetry, on a par with that of CEM.\footnote{On which of these paths should we expect to find the Everett interpretation? This is an interesting question, and although it is off the map for our present purposes (we set aside the Everett interpretation at the beginning) we stress that we do not mean to exclude the possibility that the Everett framework provides an example of how an interpretation of QM can exploit the second path.}

\section{Defending Option III}

As we observed above (\S4.1),  it is easy to argue for Option III by appealing to ordinary causal intuitions. The problem is to explain why it is \emph{those} intuitions that should be regarded as reliable, in preference to the intuition that causation acts locally -- especially if the latter intuition has physics on its side, in the sense that Option II avoids the tension between AAD and special relativity. Can the AAD Orthodoxy do better, providing an argument for Option III which does not simply appeal to intuitions? In this section we want to note two possible alternative arguments which may be found in the literature. Neither is well-developed, nor in our view especially successful, but they indicate the territory the Orthodoxy needs to explore, if it is to put its case on firmer foundations.

\subsection{The free will argument}
We noted earlier (\S2.3) that Bell's Theorem depends on two crucial assumptions, Locality and Independence. Option II turns on the proposal that Independence fails in EPRB, just as we ordinarily suppose that it fails in SEPRB. But it has been clear  since Bell's original paper  [\ref{ref:bell}] that rejection of Independence provides a \emph{formal} option for saving Locality. So why have proponents of the AAD Orthodoxy, including  Bell himself, felt able to dismiss this possibility?  

For Bell himself, and apparently for many other physicists, a crucial factor seems to have been the belief that admission of beables correlated with future measurement settings in EPRB would conflict with free will.\footnote{See [\ref{ref:bell2}, p.~244] and the remark quoted in [\ref{ref:price96}, p.~241] for Bell's view on this matter; and [\ref{ref:hooft}] for references to similar views.} The argument that these writers appear to have in mind is based on the claim that if there were already such a future-dependent beable -- call it an ``f-beable'' -- in existence, whose value was strictly correlated with some future measurement setting, that measurement setting could not now be freely chosen by an experimenter (or indeed by a random device of any other kind). If we accept this claim, and also accept that measurement settings can be freely chosen, then it follows that there are no f-beables.  

We are not aware of any writer, in physics or in philosophy, who has tried to develop this argument beyond a sketch of this kind. 
To do so would require a detailed examination of three matters, each of which would be a substantial project in its own right: (a) what is meant by ``free will'' in this context; (b) why f-beables should be incompatible with free will, in this sense; and (c) what our grounds are for confidence that there is such a thing as free will, in this sense. 
Exploring these issues would be a worthwhile project, in our view; and is an essential one, for the AAD Orthodoxy, if its argument for Option III is to rest on the claim that a violation of Independence would be in conflict with free will. 

We offer the following algorithm, in a sceptical spirit, as a way to clarify these matters. If someone claims that f-beables would be incompatible with free will, we call their attention to the so-called \emph{block universe} view, which holds that the future already exists  (inaccessible to us, in most respects, but as real as the past and the present, nevertheless). We ask them whether they regard the block universe view  as similarly incompatible with free will. If they say ``Yes'', we have reduced their objection to f-beables to a more familiar case, and gained some justification for setting it aside. Their objection to f-beables is of a piece with their objection to the block universe, and here we have many allies (including, apparently, the majority of working physicists). It may be nontrivial to say how, and in what respect, free will survives the block universe; but for this sort of opponent, the same answer will serve for f-beables, too. (Perhaps there are some senses of ``free will'' which are incompatible with the block universe; but if we have already abandoned the idea that we have free will in \textit{those} senses, they provide no further objection to f-beables.)

The other possibility is that our opponent will say that f-beables pose a new and greater threat to free will than that posed by the block universe. In this case, the algorithm recommends that we should challenge the opponent to justify the distinction between the two cases, \textit{without begging the question} -- e.g., without simply \emph{assuming} that causation can only work from past to future (which is the very question at issue), or that the past is fixed in a way which the future is not (which the block universe denies).

At present, the free will objection to f-beables seems to rest on strong intuitions but weak or nonexistent arguments. We invite those who find it persuasive to justify their intuitions, and to respond to the challenge that physics is telling us that our intuitions are mistaken.\footnote{We share the sentiment of 't Hooft [\ref{ref:hooft}], who writes: ``We dismiss all unquestioned `free will'
assumptions in physics as being not worthy of a mathematically rigorous theory.''}  For the moment, our contribution has been to put that challenge in a sharper form than has previously been available, by highlighting the choice between f-beables and fundamental asymmetries, in a case in which our intuitions seem strongly in favour of local explanation.

\subsection{Independence and consistency}

A second candidate for an argument for Option III turns on the suggestion that violations of Independence might lead to inconsistencies, or  at least to new difficulties in ensuring the coherence of the theories concerned. Here Tim Maudlin [\ref{ref:maudlin02}] deserves special mention, not only for articulating a version of such an argument, but also, more generally, as one of the very few philosophical proponents of the AAD Orthodoxy who actually offers a serious discussion of the possibility of abandoning Independence in order to preserve Locality.

Maudlin's arguments against this possibility (and hence, in our terms, in favour of Option III) are in two parts. The first part is mainly a criticism of a particular proposal due to Cramer [\ref{ref:cram80}, \ref{ref:cram86}], the so-called Transactional Interpretation of QM.\footnote{For responses to this aspect of Maudlin's argument, see [\ref{ref:berkovitz} and \ref{ref:kastner}].} The second part is more general, and turns on the idea that the kind of retrocausality introduced by a violation of Independence introduces new difficulties in arranging the global consistency of a physical theory:
\begin{quote}
If the course of present events depend on the future and the shape of the future is in part determined by the present then there must be some structure which guarantees the existence of a coherent mutual adjustment of all the free variables \ldots

\emph{Local} relativistic theories avoid these problems since solutions to the field equations at a point are constrained only by the values of quantities in one light cone (either past or future) of a point. Thus in a deterministic theory, specifying data along a hyperplane of simultaneity suffices to fix a unique solution at all times, past and future of the plane. Further, the solutions can be generated sequentially: the solution at $t=0$ can be continued to a solution at $t=1$ without having had to solve for any value at times beyond $t=1$. 
\ldots

Any theory with both backwards and forwards causation cannot have such a structure. Data along a single hypersurface do not suffice to fix the immediate future since that in turn may be affected by its own future. The metaphysical picture of the past generating the future must be abandoned, and along with it the mathematical tractability of local theories. [\ref{ref:maudlin02}, p.~201] 
\end{quote}

But Maudlin's argument sets the bar too high, in our view. By parity of reasoning, one could argue that forward causation leads to exactly the same problems: ``Data along a single hypersurface do not suffice to fix the immediate [past] since that in turn may be affected by its own [past].''\footnote{Maudlin has been misled, perhaps, by the attempt to foist an asymmetrical ``metaphysical picture of the past generating the future'' on the time-symmetric sense in which, as he himself puts it,  the data ``along a hyperplane of simultaneity suffices to fix a unique solution at all times, past and future of the plane.''} 

How is the problem solved in this case? Simply in virtue of the fact that in a deterministic theory, the data at $t=0$ fix not only the data at $t=-1$ but also the data at $t=-2$, of which the data at $t=-1$ would normally be said to be an effect. In other words, the data at $t=-2$ are not independently variable, in the sense that would give rise to trouble. Maudlin gives us no reason to think that the same solution will not work in the opposite direction, to solve the problem for retrocausality, too.\footnote{That is, the data at $t=0$ fix not only the data at $t=1$ but also the data at $t=2$, of which the data at $t=1$ would, in a retrocausal setting, be said to be an effect. Again, the data at $t=2$ are not independently variable.}

This is not to deny that there are deep puzzles in this neighbourhood. Where does our sense of the direction of causation come from, for example, in such a deterministic model? What entitles us to regard the Independence-violating kind of dependence as \emph{backward} causation? And can we properly speak of causation at all, without a kind of free will denied to us by determinism? 

These issues are important, in this context, for they seem likely to throw light on both the character and the conceptual cost of the option of abandoning Independence in order to preserve Locality. They call for far more extensive discussion than we could hope to provide in this paper.\footnote{See [\ref{ref:priwes}] and [\ref{ref:evans}] for two recent attempts to tackle these issues.} We cannot, of course, rule out the possibility that when that discussion is properly conducted, it will reveal a decisive reason for preferring Option III to Option II. Our point is simply that the jury is still out; indeed, it has hardly begun its deliberations. Until it reports, the case for the AAD Orthodoxy is seriously incomplete. 

\section{Entanglement and epistemic perspective}

We close with an analogy intended to illustrate what we take to be the lesson of the EPRB/SEPRB comparison, under the assumption that Option II turns out to be favoured. Imagine a little creature (perhaps a crab) which scans Figure 2 sideways, from left to right, with a vertical one-dimensional field of view, collecting information as it goes. Imagine in addition that its access to the properties of the particles it encounters is incomplete: it cannot observe the beables directly, but only infer them from the nature of
the interactions it encounters. When its gaze passes the mirror, and moves to the right, it knows there are two worldlines in view, one oriented upwards to the right and one oriented downwards to the right, but it cannot see the beables instantiated on either worldline.\footnote{Readers who find this example excessively metaphorical may substitute for the sideways creature a computer which collects all of the information from the spacetime region in question for a given experiment, and then delivers it to a physicist  in slices of \textit{position}, one position at a time, moving from left to right.}

Suppose that from many similar situations, our creature has learnt the nature of the correlations in such cases, between the properties of interactions such as {B} and {C}, lying further to the right. It knows, in other words, that in these circumstances the joint probability of both worldlines continuing in the same way beyond the polarizers is given by    \cscb .  How should it represent this knowledge, for the use of its own future self? In other words, how should it write down what it currently knows about the two worldlines, so as to maximise its own predictive abilities, as its gaze moves further to the right? In particular, how should it represent the way in which the probability of what it will find at {B} will depend on the polarization setting at {C}, and vice versa? The description it needs, as we see immediately from our own situation in the case of EPRB, is an entangled state, representing precisely this correlation between the ``future'' (i.e., in the imagined case, \textit{rightwards}) interactions of
one worldline and the ``future'' interactions of the other.

Our perspective is not that of this creature, of course. Our gaze crawls across the diagram from bottom to top, not from left to right. Hence it is easy for us to see that this creature's ``sideways entanglement'' is merely an epistemic state, a form of representation to which it is forced because, unlike us, it does not yet know the setting of the polarizer at {C} or {B}. But what's sauce for the goose is sauce for the gander. From this little creature's perspective, it is we whose gaze moves sideways; and we who assign a merely epistemic state to the system depicted by the pair of worldlines in EPRB (Figure 1), because we do not yet know the setting of the polarizer at {A} or {B}.

Who gets it right -- we ourselves, or this sideways creature? The interesting possibility, recommended by the evident symmetry between EPRB and SEPRB, is that both get it equally right, in so far as one can speak of right at all. There is no single, god-given, correct way to accumulate information in a lattice-like world of intersecting worldlines. We do it the way we do it because, as structures within the lattice, we have an orientation which it is not ours to change. (We do not literally crawl upwards across the lattice, of course. Rather each of us, in a timeless
sense, is merely a sequence of states within the lattice.)

If we want to know what description is \textit{really} right -- i.e., how the world is in itself -- the best we can do is to try to disentangle those elements of the descriptions we find it natural to apply to the world which reflect our own epistemic perspective, from those which belong to the world in itself, independently of our perspective. The useful feature of the symmetry between EPRB and SEPRB is that it makes this task unusually easy,
by presenting us with an alternative epistemic perspective, a mere rotation's distance from our own. By asking ourselves what varies and what is preserved under this rotation, we gain some guidance about what belongs to our perspective and what to the world in itself.

The answers suggested by the EPRB/SEPRB symmetry are certainly counterintuitive. Entanglement in EPRB becomes merely epistemic, and the underlying ontic state of the photons in EPRB comes to depend as much on conditions in the future as, in the SEPRB case, it depends on conditions to the right. But it is important to bear in mind two things. First, the proposal also explains just why we find this so counterintuitive: the intuitions bequeathed to us by our ancestors have been conditioned since first life by the temporal asymmetry of our epistemic perspective. Second, if we can free ourselves of the burden of the ancestral habits of thought, the vista that opens up is very attractive indeed. Most importantly, EPRB need no more involve any kind of spooky AAD than SEPRB does. And with AAD off the table, the second concern about nonlocality, that it conflicts with special relativity, simply melts away.

\section*{Acknowledgements}

We are grateful for comments and discussion from Nathan Argaman, Sean Carroll, Eric Cavalcanti, Richard Healey, Ruth Kastner, Matt Leifer, Gerard Milburn, David Miller, Peter Morgan, Wayne Myrvold, Travis Norsen, Giovanni Valente and Howard Wiseman, and from an anonymous referee. Huw Price acknowledges research support from the Australian Research Council and the University of Sydney. He is also grateful for stimulating discussions with participants in \emph{PIAF '09:  New Perspectives on the Quantum State,} the second annual conference of the  \emph{Perimeter Institute -- Australia Foundations Collaboration,} held at the Perimeter Institute for Theoretical Physics, Waterloo, Ontario, September 27 -- October 2, 2009.

\section*{References}

\renewcommand{\theenumi}{\arabic{enumi}}
\renewcommand{\labelenumi}{[\theenumi]}
\begin{enumerate}
\item Albert, D.Z.~\& Galchen, R., 2009. ``A quantum threat to special relativity'', \emph{Scientific American,} 300, 32--39. \label{ref:albert} 

\item Aharonov, Y., Bergmann, P.G.~\& Lebowitz, J.L., 1964. ``Time symmetry in the quantum process of measurement'', \textit{Physical Review}, 134, B1410--B1416. [\href{http://prola.aps.org/abstract/PR/v134/i6B/pB1410_1}{DOI: 10.1103/PhysRev.134.B1410}] \label{ref:aharonov}

\item Aharonov,Y. and Vaidman, L., 1991. ``Complete description of a quantum system at a given time'', {\em Journal of Physics A}, 24, 2315--2328.\\\  [\href{http://www.iop.org/EJ/abstract/0305-4470/24/10/018}{DOI: 10.1088/0305-4470/24/10/018}]

\item Argaman, N., 2008. ``On Bell's Theorem and causality''.\\\  [\href{http://arxiv.org/abs/0807.2041}{arXiv:0807.2041v1 [quant-ph]}]\label{ref:argaman}

\item Bell, J.S., 1964. ``On the Einstein Podolsky Rosen Paradox'', \emph{Physics} 1, 195--200. \label{ref:bell}

\item Bell, J.S., 2004. ``La nouvelle cuisine'', in \emph{Speakable and Unspeakable in Quantum Mechanics,} 2nd.~edition, Cambridge: Cambridge University Press, 232--248. \label{ref:bell2}

\item Berkovitz, J., 2002. ``On Causal Loops in the Quantum Realm'', in T.~Placek and J.~Butterfield (eds), {\em Non-locality and Modality}, Dordrecht: Kluwer, 235--257. \label{ref:berkovitz}

\item Blaylock, G., 2010. ``The EPR paradox, Bell's inequality, and the question of locality'', \emph{American Journal of Physics,} 78, 111--120.\\\ [\href{http://link.aip.org/link/?AJP/78/111/1}{DOI: 10.1119/1.3243279}]. \label{ref:blaylock}

\item Brukner, C., Taylor, S., Cheung, S.~\& Vedral, V., 2004. ``Quantum entanglement in time'', 
[\href{http://arxiv.org/abs/quant-ph/0402127}{arXiv:quant-ph/0402127v1}]. \label{ref:brukner}

\item Cohen, R.S., Horne, M.~\& Stachel, J. (eds), 1997. \emph{Potentiality, Entanglement and Passion-at-a-distance: Quantum Mechanical Studies for Abner Shimony, Volume Two}, Dordrecht: Kluwer Academic Publishers.\label{ref:cohen}

\item Costa de Beauregard, O., 1953. ``M\'echanique quantique'', \emph{Comptes Rendus  de l'Acad\'emie des Sciences}, T236, 1632--1634. \label{ref:costa}

\item Costa de Beauregard, O.,~1976. ``Time Symmetry and Interpretation of Quantum Mechanics'', \textit{Foundations of Physics},  6, 539--559. \\\ 
[\href{http://www.springerlink.com/content/k8l386556135g250/}{DOI: 10.1007/ BF00715107}]

\item Costa de Beauregard, O., 1977. ``Time symmetry and the Einstein paradox'', \textit{Il Nuovo Cimento}, 42B, 41--63. [\href{http://www.springerlink.com/content/x755748406278g65/}{DOI: 10.1007/BF02906749}]

\item Cramer, J.G., 1980. ``Generalized absorber theory and the Einstein-Podolsky-Rosen paradox'', \textit{Physical Review D}, 22, 362--376. \\\hspace{0pt}
[\href{http://prd.aps.org/abstract/PRD/v22/i2/p362_1}{DOI: 10.1103/ PhysRevD.22.362}]\label{ref:cram80}

\item Cramer, J.G., 1986: ``The transactional interpretation of quantum mechanics'',  \textit{Reviews of Modern Physics}, 58, 647--687. \\\ [\href{http://rmp.aps.org/abstract/RMP/v58/i3/p647_1}{DOI: 10.1103/RevModPhys.58.647}]\label{ref:cram86}

\item Evans, P.W., 2010. ``Retrocausality at no extra cost''. \\\ \href{http://philsci-archive.pitt.edu/archive/00005361/}{http://philsci-archive.pitt.edu/archive/00005361/}.\label{ref:evans}

\item Fuchs, C.A.~\& Peres, A., 2000. ``Quantum Theory Needs No `Interpretation{'}'',  \textit{Physics Today}, 53, 70--71.\label{ref:fuchs}

\item Hokkyo, N., 1988. ``Variational formulation of transactional and related interpretations of quantum mechanics'', \emph{Foundations of Physics Letters,} 1, 293--299. [\href{http://www.springerlink.com/content/p81r939075727216/}{DOI: 10.1007/BF00690070}]\label{ref:hokk}

\item Kastner, R., 2006. ``Cramer's Transactional Interpretation and Causal Loop Problems'',  {\em Synthese} 150, 1--14. [\href{http://www.springerlink.com/content/t701157ru1802344/}{DOI: 10.1007/s11229-004-6264-9}] \label{ref:kastner}

\item Leifer, M.S., 2006. ``Quantum dynamics as an analog of conditional probability'', \emph{Phys.~Rev.~A} 74, 042310. [\href{http://link.aps.org/doi/10.1103/PhysRevA.74.042310}{DOI: 10.1103/PhysRevA.74.042310}]  [\href{http://arxiv.org/abs/quant-ph/0606022}{arXiv:quant-ph/0606022v2}]\label{ref:leif06}

 \item Maudlin, T., 2002. \textit{Quantum Non-Locality and Relativity}, 2nd edn., Oxford: Blackwell Publishing. \label{ref:maudlin02}

\item Maudlin, T., 2010. ``What Bell proved: A reply to Blaylock'', \emph{American Journal of Physics,} 78, 121--125. [\href{http://link.aip.org/link/?AJP/78/121/1}{DOI: 10.1119/1.3243280}]. \label{ref:maudlin}

\item Miller, D.J., 1996. ``Realism and time symmetry in quantum mechanics'', \emph{Physics Letters,} A222, 31--36. [\href{http://linkinghub.elsevier.com/retrieve/pii/0375960196006202}{DOI: 10.1016/0375-9601(96)00620-2}]\label{ref:mill96}

\item Miller, D.J., 1997. ``Conditional probabilities in quantum mechanics from a time-symmetric formulation'', \emph{Il Nuovo Cimento,} 112B, 1577--1592.\label{ref:mill97}

\item Norsen, T., 2009. ``Local Causality and Completeness: Bell vs. Jarrett'',  {\em Foundations of Physics} 39, 273--294. [\href{http://www.springerlink.com/content/d44771273592530r/}{DOI: 10.1007/s10701-009-9281-1}] \label{ref:norsen}

\item Peskin, M. and Schroeder, D., 1995. \emph{An Introduction to Quantum Field Theory,} Westview Press.

\item Price, H., 1984. ``The philosophy and physics of affecting the past'', \textit{Synthese}, 61, 299--324. [\href{http://www.springerlink.com/content/p94624888311jv41/}{DOI: 10.1007/BF00485056}]\label{ref:price84}

\item Price, H., 1994. ``A neglected route to realism about quantum mechanics'',  \textit{Mind}, 103, 303--336. [\href{http://arxiv.org/abs/gr-qc/9406028}{arXiv:gr-qc/9406028v1}]

\item Price, H., 1996. \textit{Time's Arrow and Archimedes' Point}, New York: Oxford University Press.\label{ref:price96}

\item Price, H., 1997. ``Time symmetry in microphysics'',  \emph{Philosophy of Science,} 64, S235--244. [\href{http://arxiv.org/abs/quant-ph/9610036}{arXiv:quant-ph/9610036v1}] \label{ref:price97}

\item Price, H., 2001. ``Backward causation, hidden variables, and the meaning of completeness'', \textit{PRAMANA -- Journal of Physics}, 56, 199--209.\\\ [\href{http://www.springerlink.com/content/343t87014l507086/}{DOI: 10.1007/s12043-001-0117-6}]

\item Price, H., 2008. ``Toy models for retrocausality'', \emph{Studies in History and Philosophy of Mod.~Physics,} 39, 752--76. [\href{http://arxiv.org/abs/0802.3230}{arXiv:0802.3230v1 [quant-ph]}].\label{ref:price08}  

\item Price, H., 2010. ``Time-symmetry without retrocausality: how the quantum can withhold the solace''.  [\href{http://arxiv.org/abs/1002.0906}{arXiv:1002.0906v1 [quant-ph]}]\label{ref:price10}

 \item Price, H. and Weslake, B., 2010. ``The Time-Asymmetry of Causation'', in H.~Beebee, C.~Hitchcock and P.~Menzies (eds), {\em The Oxford Handbook of Causation}, New York: Oxford University Press, pp.~414--443.\label{ref:priwes}

\item Rietdijk, C.W., 1978. ``Proof of a retroactive influence'', \textit{Foundations of Physics}, 8,  615--628. [\href{http://www.springerlink.com/content/w445263507151566/}{DOI: 10.1007/BF00717585}]\label{ref:reit}

\item Shimony, A. 1984. ``Controllable and uncontrollable non-locality'', in S.~Kamefuchi \emph{et al.}~(eds), \emph{Foundations of Quantum Mechanics in Light of the New Technology,} Tokyo: Physical Society of Japan, 225--230.\label{ref:shim}

\item Sutherland, R.I., 1983. ``Bell's theorem and backwards-in-time causality'', \textit{International Journal of Theoretical Physics}, 22, 377--384. \\\ [\href{http://www.springerlink.com/content/k8g22512n5w38xu1/}{DOI: 10.1007/BF02082904}]\label{ref:sutherland83}

\item Sutherland, R.I., 1998. ``Density formalism for quantum theory'', \emph{Foundations of Physics.} 28, 1157--1190. [\href{http://www.springerlink.com/content/u551710rp1x27723/}{DOI: 0.1023/A:1018850120826}]

\item Sutherland, R.I., 2008. ``Causally symmetric Bohm model'', \emph{Studies in History and Philosophy of Modern Physics,} 39, 782--805. \\\ [\href{http://arxiv.org/abs/quant-ph/0601095}{arXiv:quant-ph/0601095v2}] \label{ref:sutherland}

\item 't Hooft, G., 2007. ``The free-will postulate in quantum mechanics'.  \\\ [\href{http://arxiv.org/abs/arXiv:quant-ph/0701097}{arXiv: quant-ph/0701097v1}]. \label{ref:hooft}

\item Wharton, K.B., 2007. ``Time-symmetric quantum mechanics'', {\em Foundations of  Physics,} 37, 159--168. [\href{http://www.springerlink.com/content/v765h37416011n63/?p=877db523fa624fd582e512cfb9a5bebb&pi=1}{DOI: 10.1007/s10701-006-9089-1}] \label{ref:wharton07}

\item Wharton, K.B., 2009. ``A novel interpretation of the Klein-Gordon equation'', \emph{Foundations of Physics,} 40, 313--332. \\\ [\href{http://www.springerlink.com/content/l14jr4x24pk82521/}{DOI: 10.1007/s10701-009-9398-2}]  [\href{http://arxiv.org/abs/0706.4075}{arXiv:0706.4075v3 [quant-ph]}].
\label{ref:wharton09}

\end{enumerate}
\end{document}